\documentclass[journal,twoside,web]{ieeecolor}
\usepackage{generic}
\usepackage{cite}
\usepackage{amsmath,amssymb,amsfonts}
\usepackage{graphicx}
\usepackage{booktabs}
\usepackage{multirow}
\usepackage{hyperref}
\hypersetup{hidelinks=true}
\usepackage{textcomp}
\def\BibTeX{{\rm B\kern-.05em{\sc i\kern-.025em b}\kern-.08em
    T\kern-.1667em\lower.7ex\hbox{E}\kern-.125emX}}
\begin{document}

\title{CardioFusion-AI: Robust ECG--PPG Fusion for Multimodal Physiological Monitoring Under Signal Degradation}

\author{Navaneeth~Krishnan~K and Janakiraman~K
\thanks{Manuscript submitted for review.}
\thanks{N. Krishnan K. is with the Department of Electronics and Communication Engineering, Saveetha School of Engineering ( e-mail: knavaneeth385@gmail.com) and J. K is with the Government Kilpauk Medical College and Hospital ( e-mail: kjanaki2408@gmail.com).}}

\maketitle

\begin{abstract}
Wearable electrocardiogram (ECG) and photoplethysmogram (PPG) sensors are complementary but individually fragile: motion artifact, poor contact, and sensor dropout routinely degrade one or both signals. Fusion strategies that assume both modalities are equally trustworthy can become \emph{less} reliable than a single clean modality once degradation sets in, yet the comparative robustness of different ECG--PPG fusion mechanisms under graded degradation and complete modality loss remains poorly characterized. We present CardioFusion-AI, a framework whose signal-processing front end (R-peak and systolic-peak detection, an Orphanidou-type signal-quality index, and beat-by-beat pulse transit time estimation) is validated on 53 real intensive-care recordings (848 windows, heart-rate mean absolute error 1.61 bpm ECG / 2.78 bpm PPG) and a real annotated fetal ECG database (R-peak F1 0.89--0.98). Building on this validated front end, we conduct a controlled, physiologically-grounded synthetic degradation study comparing eight ECG--PPG fusion strategies -- unimodal baselines, three fixed-weight fusion variants, attention fusion, and two sample-adaptive gated fusion mechanisms (one conditioned only on learned embeddings, one additionally conditioned on the validated signal-quality index) -- across six degradation regimes spanning graded corruption and complete modality loss, over five independent training seeds. Attention fusion achieved the lowest descriptive overall error (1.66$\pm$0.43 bpm). Both adaptive gates correctly reallocated weight toward the healthy modality under complete modality loss but showed near-zero correlation between gate weight and signal quality when both modalities were present but graded-degraded ($r\approx0.10$--$0.24$), and signal-quality conditioning produced a large, specific improvement under missing-PPG conditions (1.56$\pm$0.59 bpm, near the 1.48 bpm unimodal ceiling). With only five training seeds, no pairwise comparison survives Holm-corrected significance testing; we report effect sizes and confidence intervals accordingly. These results indicate that modality availability and modality quality are functionally distinct problems for adaptive fusion.
\end{abstract}


\section{Introduction}
\label{sec:introduction}

\IEEEPARstart{W}{earable} physiological monitoring increasingly relies on multiple complementary biosignals rather than any single sensor. Electrocardiography (ECG) and photoplethysmography (PPG) are the most common pairing in consumer and clinical wearables: ECG provides a direct electrical readout of cardiac timing, while PPG provides an optically derived pulse waveform from which heart rate, pulse timing, and -- in combination with ECG -- pulse transit time (PTT) can be estimated \cite{mukkamala2015}. Because the two modalities are captured through different physical transduction mechanisms, they are also affected by different degradation processes, which is a large part of why fusing them is attractive: a corruption that disables one modality frequently leaves the other intact.

In practice, however, both modalities degrade routinely and often simultaneously. Motion artifact corrupts PPG far more readily than ECG; poor electrode contact or perspiration corrupts ECG independently of PPG; and either sensor can lose contact entirely for extended periods in ambulatory use. A fusion method that implicitly assumes both inputs are equally trustworthy can therefore become \emph{less} reliable than a single well-behaved modality once one input deteriorates -- naive averaging of a clean signal with a corrupted one degrades the estimate relative to simply discarding the corrupted input.

This motivates \emph{signal-quality-aware} fusion: mechanisms that adjust how much each modality contributes based on how trustworthy it currently is. Several families of fusion strategy have been proposed for multimodal biosignal and, more broadly, multimodal machine learning problems, including early/input-level concatenation, feature-level fusion via a learned combination MLP, late/decision-level fusion with fixed or learned per-modality weights, cross-modal attention, and sample-adaptive gating in which a small network learns to reweight modalities on a per-input basis \cite{baltrusaitis2019}. Adaptive gating is conceptually the most attractive of these for degraded wearable sensing, since it promises to reweight modalities exactly when and where degradation occurs, rather than applying a single global weighting scheme.

Existing evaluations of these fusion mechanisms are, however, typically conducted under clean or narrowly controlled conditions, or evaluate a single proposed mechanism against a small number of baselines. The comparative behavior of a broad set of fusion mechanisms -- including whether an adaptive, signal-quality-conditioned gate actually learns to track \emph{graded} signal degradation, as opposed to only the binary presence or absence of a modality -- has not, to our knowledge, been systematically characterized under a controlled degradation protocol with matched architectures, matched training procedures, and an explicit accounting for the statistical power of the comparison.

In this work we address that gap directly. We build CardioFusion-AI, a two-part evaluation: (i) a signal-processing front end -- peak detection, an Orphanidou-type signal-quality index (SQI) \cite{orphanidou2015}, and beat-by-beat PTT estimation -- validated against real, independently sourced physiological data; and (ii) a controlled, physiologically-grounded synthetic degradation study in which eight ECG--PPG fusion architectures, sharing an identical encoder backbone and training protocol, are compared across six degradation regimes spanning graded single- and dual-modality corruption and complete modality loss. We deliberately frame this work in terms of \emph{physiological monitoring robustness} rather than disease diagnosis or clinical risk stratification: the evidence we present concerns the reliability of physiological estimation under signal degradation, not diagnostic or prognostic performance, and should not be read as either.

The central hypothesis motivating this study is:
\begin{quote}
Signal-quality-aware and multimodal fusion mechanisms can improve the robustness of physiological estimates under ECG and/or PPG signal degradation, compared with fixed or unimodal estimation strategies; and sample-adaptive fusion guided explicitly by signal-quality information will show greater robustness to modality-specific degradation than fusion strategies without such conditioning.
\end{quote}
As we show, the data support the first clause but only partially support the second, in a way that is itself informative about the mechanism.

The contributions of this paper are threefold:
\begin{enumerate}
\item A signal-quality-validated ECG--PPG fusion framework and controlled degradation-and-modality-loss evaluation protocol for physiological estimation, built on physiologically-grounded synthetic signal generation with independently controllable, graded per-modality corruption.
\item A systematic, matched-protocol comparison of eight unimodal and multimodal fusion strategies -- including fixed-weight, attention-based, and implicit- and explicit-signal-quality-conditioned adaptive gating -- across six degradation regimes and five independent training replicates, with pre-registered, multiplicity-corrected statistical analysis.
\item A mechanistic analysis of learned modality allocation showing that adaptive gates reliably detect complete modality absence but exhibit near-zero correlation with graded signal quality when both modalities are present, and that explicit signal-quality conditioning improves this behavior selectively rather than uniformly.
\end{enumerate}

\section{Methods}
\label{sec:methods}

\subsection{Overview and Two-Part Evaluation Strategy}
CardioFusion-AI combines two categories of evidence that are epistemically distinct and are kept explicitly separate throughout this paper. \emph{Claim A} concerns the signal-processing front end -- peak detection, SQI computation, and PTT estimation -- and is supported by validation against real, third-party physiological recordings (Section~\ref{sec:realdata}). \emph{Claim B} concerns the comparative robustness of fusion \emph{architectures} under degradation, which we study using a controlled synthetic protocol (Sections~\ref{sec:degradation}--\ref{sec:stats}) because no public dataset provides graded, independently-controllable, ground-truth-labeled degradation of both ECG and PPG simultaneously. Claim B numbers should not be interpreted as clinical performance figures; they establish comparative, mechanistic behavior among fusion strategies under a controlled corruption model.

\subsection{Real-Data Validation of the Signal-Processing Pipeline}
\label{sec:realdata}
The peak-detection and signal-quality components used throughout this work were validated, independently of the degradation study, against two public PhysioNet resources \cite{goldberger2000}.

\emph{R-peak detection.} Using the Abdominal and Direct Fetal ECG Database \cite{jezewski2012} (five records, cardiologist-verified beat annotations), direct-lead R-peak detection achieved an F1 score of 0.89--0.98 (mean 0.95) against expert annotation. Applying the same detector naively to the far noisier abdominal leads, without any signal-quality gating, collapsed F1 to 0.19--0.34, and per-record detection F1 correlated strongly with per-record SQI pass rate (Pearson $r=0.927$), directly motivating quality-aware processing.

\emph{Heart-rate and pulse-rate agreement.} Using the BIDMC PPG and Respiration Database \cite{pimentel2017} (53 real ICU patients, 30 s windows, 848 windows total), ECG-derived heart rate agreed with the bedside monitor reference with mean absolute error (MAE) 1.61 bpm (correlation 0.911) and PPG-derived pulse rate with MAE 2.78 bpm (correlation 0.879).

\emph{Pulse transit time.} A beat-by-beat PTT estimator, validated on the same 53-subject BIDMC cohort (7,868 individually matched beats after fixing a boundary-artifact bug in an earlier whole-window cross-correlation approach), produced a physiologically plausible distribution (mean 113.9 ms, median 104.0 ms, SD 55.6 ms) consistent with a critically-ill ICU population.

These results establish that the front-end signal processing -- and specifically the SQI computation used as an input to the signal-quality-conditioned fusion gate in Section~\ref{sec:architectures} -- is grounded in real, independently verifiable performance, even though the fusion-degradation comparison itself (Claim B) uses controlled synthetic data.

\subsection{Controlled Synthetic Degradation Study: Signal Generation}
\label{sec:degradation}
No public dataset provides simultaneous ECG and PPG with independently controllable, graded degradation severity and known ground-truth physiological state; real degraded recordings confound degradation type, severity, and physiological state in ways that make controlled comparison of fusion mechanisms impossible. We therefore generated physiologically-grounded synthetic ECG and PPG using NeuroKit2 \cite{makowski2021}, whose ECG simulator implements the McSharry dynamical model \cite{mcsharry2003} and whose PPG simulator includes built-in, literature-standard motion-artifact, baseline-drift, and burst-noise parameters.

For each 8 s window (sampling rate 125 Hz, 1000 samples), a target heart rate was drawn as $\mathrm{HR}\sim\mathcal{U}(50,110)$ bpm and used to generate a clean ECG and a clean PPG waveform sharing that heart rate. Each modality was then independently corrupted to one of five severity levels: \emph{clean} (0), \emph{mild} (1), \emph{moderate} (2), \emph{severe} (3), or \emph{missing} (4, complete sensor dropout modeled as a noise floor with no cardiac structure). ECG degradation combined additive Gaussian noise, low-frequency baseline wander, and randomly-timed motion bursts, with severity-scaled amplitudes; PPG degradation used NeuroKit2's native motion-amplitude, powerline, drift, and burst-noise parameters, also severity-scaled. Table~\ref{tab:degradation} summarizes the resulting evaluation grid, which is the full factorial of graded co-degradation (16 combinations, levels 0--3 $\times$ 0--3) plus two targeted complete-modality-loss conditions, collapsing to six analysis regimes.

\begin{table}[t]
\centering
\caption{Controlled degradation conditions. ``Level'' spans clean (0) through severe (3); ``missing'' (4) denotes complete sensor dropout and is evaluated only against a clean partner modality.}
\label{tab:degradation}
\begin{tabular}{@{}lccl@{}}
\toprule
Regime & ECG level & PPG level & \# grid cells \\
\midrule
Both clean & 0 & 0 & 1 \\
ECG degraded only & 1--3 & 0 & 3 \\
PPG degraded only & 0 & 1--3 & 3 \\
Both degraded & 1--3 & 1--3 & 9 \\
ECG missing & 4 & 0 & 1 \\
PPG missing & 0 & 4 & 1 \\
\bottomrule
\end{tabular}
\end{table}

Windows are i.i.d.\ synthetic draws with no subject identity, so there is no possibility of subject-level leakage across splits; independence across the train/validation/test partition instead follows directly from using disjoint random-seed ranges for signal generation in each split (Section~\ref{sec:protocol}). Dataset sizes (792 train / 234 validation / 324 test window pairs, approximately 44/13/18 windows per grid cell) were chosen to keep the eight-architecture $\times$ five-seed sweep computationally tractable on a single CPU core while providing at least $\sim$20 test windows per regime cell for stable per-regime error estimates; we report this as a scale limitation in Section~\ref{sec:limitations} rather than presenting it as an optimized choice.

The regression target is the true, generator-known heart rate (bpm) for the window, which -- unlike reference values derived from noisy real recordings -- provides an exact ground truth against which fusion-strategy accuracy can be compared without reference-signal error confounding the comparison.

\subsection{Signal Quality Assessment Applied to Synthetic Windows}
The SQI used both diagnostically and as an input feature to the signal-quality-conditioned fusion gate is the same Orphanidou-type index validated in Section~\ref{sec:realdata} \cite{orphanidou2015}: beat detection followed by a template-correlation-based quality score and a feasibility flag (physiologically plausible rate and beat-to-beat regularity). We emphasize that this SQI is a signal-quality \emph{descriptor}, derived from waveform morphology and rate regularity, not a ground-truth quality label; it can and does fail to discriminate in some conditions (Section~\ref{sec:discussion}). Applied to the synthetic test windows, mean SQI decreased monotonically with degradation severity for both modalities (ECG: 0.988 at level 0 to 0.946 at level 3; PPG: 0.995 to 0.952) and collapsed sharply under complete dropout (ECG missing: mean SQI 0.19, feasibility-pass rate 27\%; PPG missing: mean SQI 0.05, feasibility-pass rate 14\%), confirming the SQI is directionally informative before it is used as a gate input.

\subsection{Fusion Architectures}
\label{sec:architectures}
\begin{figure}[t]
\centering
\includegraphics[width=\linewidth]{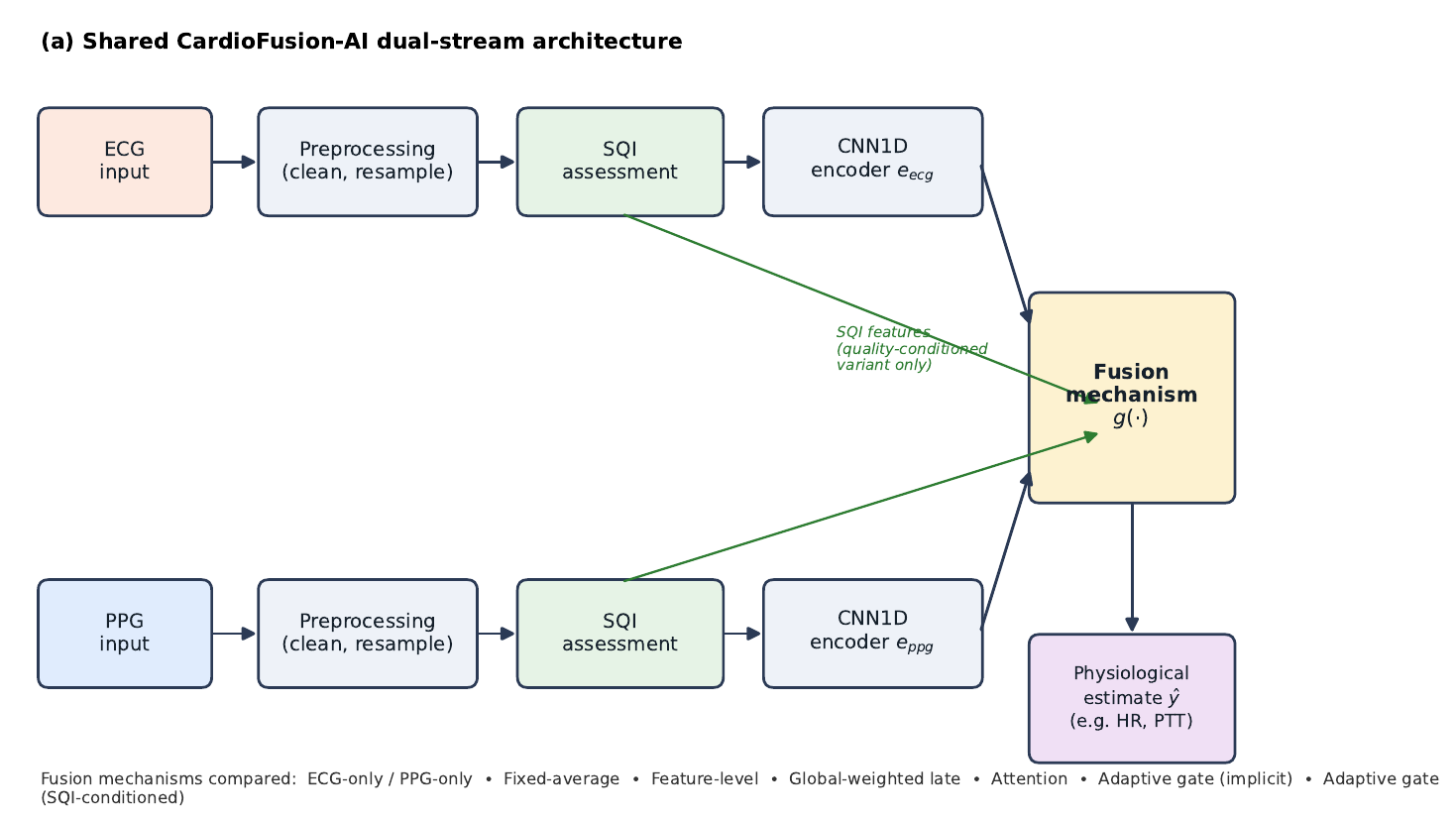}
\caption{Shared CardioFusion-AI dual-stream architecture. Each modality is independently preprocessed, quality-assessed, and encoded; the fusion mechanism (the only component that differs across the eight compared strategies) combines the two embeddings into a physiological estimate. The signal-quality index is an explicit gate input only in the SQI-conditioned adaptive variant.}
\label{fig:architecture}
\end{figure}

Fig.~\ref{fig:architecture} shows the shared dual-stream architecture common to all eight strategies; they differ only in the fusion mechanism $g(\cdot)$ that combines the two per-modality embeddings.

All eight architectures share an identical single-modality encoder backbone: a four-block 1-D CNN (kernel size 7, channel progression 16--32--64--64, batch normalization, ReLU, dropout 0.2, max-pooling after each block) followed by global average pooling and a linear projection to a $d=64$-dimensional embedding, $e_{\mathrm{ecg}}=f_{\theta_{\mathrm{ecg}}}(x_{\mathrm{ecg}})$, $e_{\mathrm{ppg}}=f_{\theta_{\mathrm{ppg}}}(x_{\mathrm{ppg}})\in\mathbb{R}^{64}$. Holding the encoder architecture fixed isolates the fusion mechanism as the only varying factor. A regression head $h:\mathbb{R}^{64}\to\mathbb{R}$ (two-layer MLP, ReLU, dropout 0.3) maps the fused representation to a scalar heart-rate estimate $\hat y$.

\begin{enumerate}
\item \textbf{ECG-only} / \textbf{PPG-only}: $\hat y = h(e_{\mathrm{ecg}})$ or $h(e_{\mathrm{ppg}})$; no fusion.
\item \textbf{Fixed-average fusion}: $\hat y = h\!\left(\tfrac{1}{2}e_{\mathrm{ecg}}+\tfrac{1}{2}e_{\mathrm{ppg}}\right)$, a static, input-independent 50/50 combination -- the most literal realization of ``fixed fusion.''
\item \textbf{Feature-level fusion}: $\hat y = h\!\left(\mathrm{MLP}([e_{\mathrm{ecg}};e_{\mathrm{ppg}}])\right)$, where the fusion MLP is learned during training but, once trained, applies the same transformation to every input regardless of quality.
\item \textbf{Global-weighted late fusion}: independent per-modality regression heads $h_{\mathrm{ecg}}, h_{\mathrm{ppg}}$ produce $\hat y_{\mathrm{ecg}}=h_{\mathrm{ecg}}(e_{\mathrm{ecg}})$, $\hat y_{\mathrm{ppg}}=h_{\mathrm{ppg}}(e_{\mathrm{ppg}})$, combined as $\hat y = w_1\hat y_{\mathrm{ecg}} + w_2\hat y_{\mathrm{ppg}}$ with $w=\mathrm{softmax}(\phi)$, $\phi\in\mathbb{R}^2$ a single learned parameter pair shared across \emph{all} samples (fixed at inference).
\item \textbf{Attention fusion}: bidirectional multi-head cross-attention \cite{vaswani2017} (4 heads) between the pooled embeddings, treated as length-1 token sequences: $\tilde e_{\mathrm{ecg}} = \mathrm{LN}(e_{\mathrm{ecg}} + \mathrm{MHA}(e_{\mathrm{ecg}}, e_{\mathrm{ppg}}, e_{\mathrm{ppg}}))$ and symmetrically for $\tilde e_{\mathrm{ppg}}$; $\hat y = h([\tilde e_{\mathrm{ecg}};\tilde e_{\mathrm{ppg}}])$. Attention weights are input-dependent but are not explicitly supervised by any signal-quality signal.
\item \textbf{Adaptive gate (implicit)}: a gate network $g=\mathrm{softmax}(\mathrm{MLP}([e_{\mathrm{ecg}};e_{\mathrm{ppg}}]))\in\Delta^1$ produces a \emph{per-sample} weight pair from the embeddings alone; $\hat y = h(g_1 e_{\mathrm{ecg}} + g_2 e_{\mathrm{ppg}})$.
\item \textbf{Adaptive gate (SQI-conditioned)}: identical to (6) except the gate additionally receives the real SQI descriptor $s=[\mathrm{SQI}_{\mathrm{ecg}},\,\mathrm{feas}_{\mathrm{ecg}},\,\mathrm{SQI}_{\mathrm{ppg}},\,\mathrm{feas}_{\mathrm{ppg}}]\in\mathbb{R}^4$ from Section~\ref{sec:realdata}: $g=\mathrm{softmax}(\mathrm{MLP}([e_{\mathrm{ecg}};e_{\mathrm{ppg}};s]))$.
\end{enumerate}
Architectures 3, 5, 6, and 7 reuse the CardioFusion-AI repository's own implementations unmodified (with the classification head reduced to a single output unit for regression); architectures 2 and 4 were added for this study as literal fixed-fusion comparators, since no fusion strategy in the base repository was fully input-independent at inference.

\subsection{Experimental Protocol}
\label{sec:protocol}
Each of the eight architectures was trained independently under five random seeds (0--4), which govern both parameter initialization and minibatch ordering; the train/validation/test partition itself was fixed and shared across all seeds and architectures, generated once using disjoint random-number-generator seed ranges per split (Section~\ref{sec:degradation}). All models were trained with Adam (learning rate $5\times10^{-4}$, weight decay $10^{-5}$, gradient-norm clipping at 1.0 -- both introduced after an initial pilot run diverged for one architecture/seed combination, then applied uniformly to every architecture for fairness), batch size 32, mean-squared-error loss, up to 30 epochs with early stopping (patience 6) on validation MSE. Evaluation used the single fixed test partition (324 window pairs) for every architecture and seed.

\subsection{Statistical Analysis}
\label{sec:stats}
For each architecture and regime we report the mean and standard deviation of test MAE across the five seeds. For a pre-registered family of three architecture comparisons (SQI-conditioned vs.\ implicit adaptive gate; SQI-conditioned adaptive gate vs.\ attention; implicit adaptive gate vs.\ attention), evaluated across seven metrics (overall MAE and the six regime-specific MAEs; 21 tests total), we treat the \emph{seed} -- one independently initialized and trained model instance -- as the unit of statistical inference, since the test partition is identical across seeds and pooling seed $\times$ window pairs as if independent would be pseudo-replication and would invalidate resulting $p$-values. We report the paired difference in per-seed mean MAE ($n=5$), an exact paired Wilcoxon signed-rank test, a paired $t$-test, a paired effect size (Cohen's $d_z$), and a $t$-based 95\% confidence interval on the mean difference, explicitly caveated given $n=5$. Holm--Bonferroni correction is applied across the full 21-test family. We separately report the correlation between SQI and learned gate weight (Section~\ref{sec:mechanistic}), computed per seed and averaged, to avoid the same pseudo-replication problem.

\section{Results}
\label{sec:results}

\subsection{Overall Fusion Performance}
Table~\ref{tab:mainresults} reports test MAE (mean $\pm$ SD across five seeds) for all eight architectures across the six degradation regimes. Attention fusion exhibited the lowest descriptive overall MAE (1.66 $\pm$ 0.43 bpm), followed by the implicit (2.06 $\pm$ 0.51 bpm) and SQI-conditioned (1.99 $\pm$ 0.61 bpm) adaptive gates; fixed-weight fusion variants (fixed-average, feature-level, global-weighted late) and the single-modality baselines were descriptively worse (2.18--2.87 bpm). We report these as descriptive comparisons: as detailed in Section~\ref{sec:stats-results}, none of the pairwise comparisons among attention fusion and the two adaptive gates survives Holm-corrected significance testing at $n=5$ seeds.

\begin{table*}[t]
\centering
\caption{Test-set MAE (bpm), mean $\pm$ SD across 5 independent training seeds, by degradation regime.}
\label{tab:mainresults}
\begin{tabular}{@{}lccccccc@{}}
\toprule
Model & Overall & Both clean & ECG degraded & PPG degraded & Both degraded & ECG missing & PPG missing \\
\midrule
ECG-only & 2.18$\pm$0.45 & 1.86$\pm$0.76 & 1.52$\pm$0.41 & 1.41$\pm$0.56 & 1.43$\pm$0.41 & 14.34$\pm$0.38 & 1.48$\pm$0.67 \\
PPG-only & 2.61$\pm$0.30 & 1.49$\pm$0.32 & 1.38$\pm$0.30 & 1.92$\pm$0.22 & 2.13$\pm$0.35 & 1.47$\pm$0.34 & 15.02$\pm$1.51 \\
Fixed-average fusion & 2.63$\pm$0.64 & 2.39$\pm$0.71 & 2.11$\pm$0.89 & 2.14$\pm$0.73 & 2.17$\pm$0.76 & 7.01$\pm$0.82 & 5.77$\pm$1.33 \\
Feature-level fusion & 2.87$\pm$0.53 & 2.53$\pm$0.57 & 2.17$\pm$0.31 & 2.25$\pm$0.53 & 2.40$\pm$0.67 & 6.95$\pm$0.76 & 7.28$\pm$1.22 \\
\textbf{Attention fusion} & \textbf{1.66$\pm$0.43} & \textbf{1.24$\pm$0.26} & 1.65$\pm$0.39 & \textbf{1.18$\pm$0.18} & \textbf{1.55$\pm$0.31} & 3.17$\pm$1.91 & 2.97$\pm$1.73 \\
Global-weighted late fusion & 2.67$\pm$0.30 & 1.98$\pm$0.44 & 1.83$\pm$0.62 & 2.05$\pm$0.35 & 2.26$\pm$0.40 & 6.86$\pm$0.54 & 7.24$\pm$0.80 \\
Adaptive gate (implicit) & 2.06$\pm$0.51 & 1.52$\pm$0.36 & 1.73$\pm$0.16 & 1.72$\pm$0.44 & 1.77$\pm$0.40 & 5.38$\pm$4.11 & 3.85$\pm$5.54 \\
Adaptive gate (SQI-cond.) & 1.99$\pm$0.61 & 1.98$\pm$0.48 & 1.88$\pm$0.69 & 1.81$\pm$0.55 & 1.87$\pm$0.68 & 4.36$\pm$1.59 & \textbf{1.56$\pm$0.59} \\
\bottomrule
\end{tabular}
\end{table*}

\subsection{Performance Across Degradation Severity}
\begin{figure}[t]
\centering
\includegraphics[width=\linewidth]{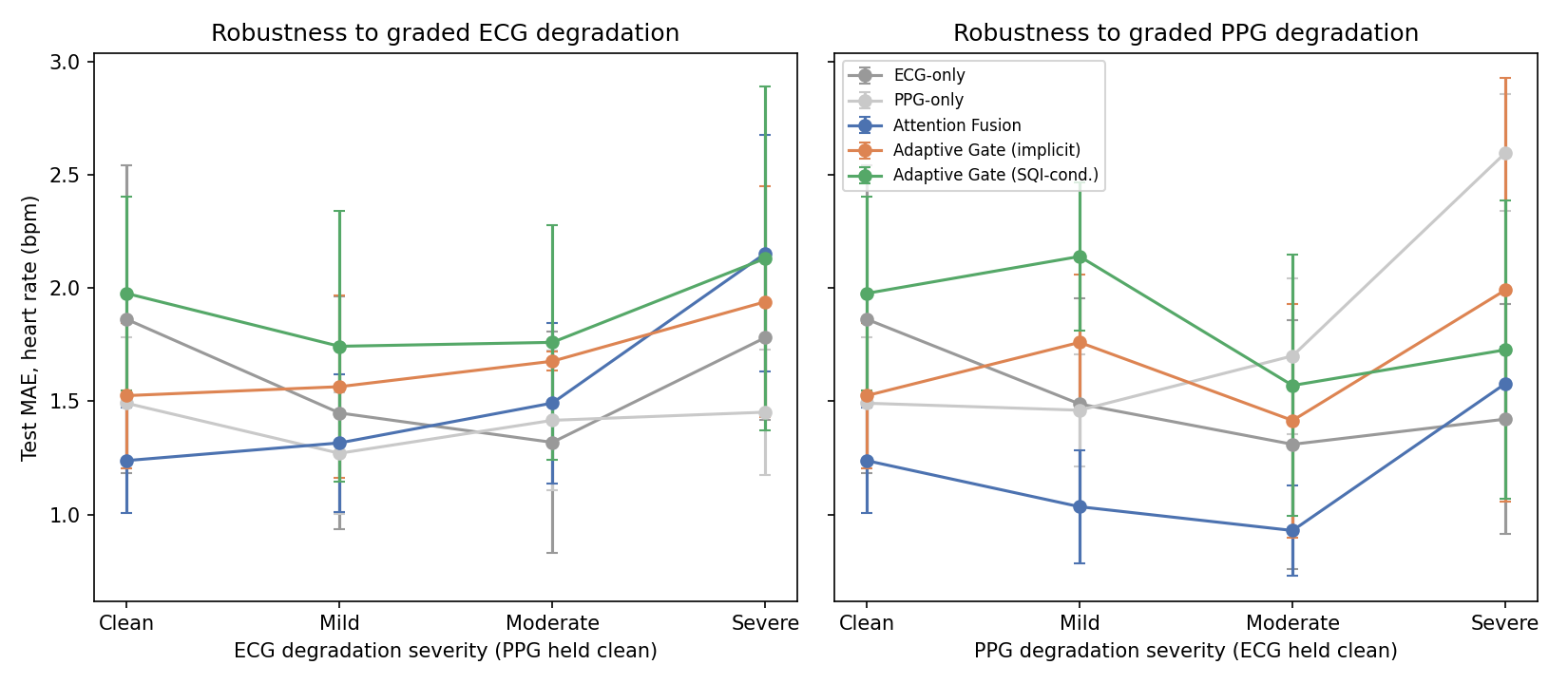}
\caption{Test MAE vs.\ graded degradation severity (0 = clean to 3 = severe), with the partner modality held clean. Error bars: SD across 5 seeds.}
\label{fig:severity}
\end{figure}

Fig.~\ref{fig:severity} shows test MAE as a function of graded degradation severity, isolated separately for ECG degradation (PPG held clean) and PPG degradation (ECG held clean). Performance trajectories differed markedly by degradation type and fusion mechanism rather than degrading uniformly. Under graded PPG degradation, attention fusion remained the strongest strategy at every severity level. Under graded \emph{ECG} degradation, by contrast, attention fusion's advantage eroded with increasing severity and reversed at the severe level, where a naive PPG-only estimator outperformed it (1.45 vs.\ 2.15 bpm) -- attention fusion's overall advantage in Table~\ref{tab:mainresults} is therefore not uniform across degradation types, and severe single-modality ECG degradation is a specific failure mode for the otherwise-strongest architecture.

\subsection{Robustness to Complete Modality Loss}
Complete modality loss (Table~\ref{tab:mainresults}, rightmost two columns) separated the architectures far more sharply than graded degradation. Single-modality baselines behaved exactly as expected -- catastrophic when their own modality is missing (ECG-only: 14.34 bpm with ECG missing; PPG-only: 15.02 bpm with PPG missing) and near-ceiling when the \emph{other} modality is missing (ECG-only: 1.48 bpm with PPG missing; PPG-only: 1.47 bpm with ECG missing), which we treat as an approximate empirical ceiling for what any fusion strategy relying on the surviving modality alone could achieve.

Fixed-weight fusion strategies (fixed-average, feature-level, global-weighted late) could not approach this ceiling under either missing-modality condition (5.77--7.28 bpm), since none can reduce a dead channel's contribution at inference. The SQI-conditioned adaptive gate came within 0.08 bpm of the unimodal ceiling under missing PPG (1.56 $\pm$ 0.59 bpm vs.\ 1.48 bpm), while under missing ECG it reached 4.36 $\pm$ 1.59 bpm -- better than the implicit gate (5.38 $\pm$ 4.11 bpm) and the fixed-weight strategies, but well short of the corresponding 1.47 bpm ceiling. This ECG/PPG asymmetry is examined mechanistically in Section~\ref{sec:mechanistic}.

\begin{figure}[t]
\centering
\includegraphics[width=\linewidth]{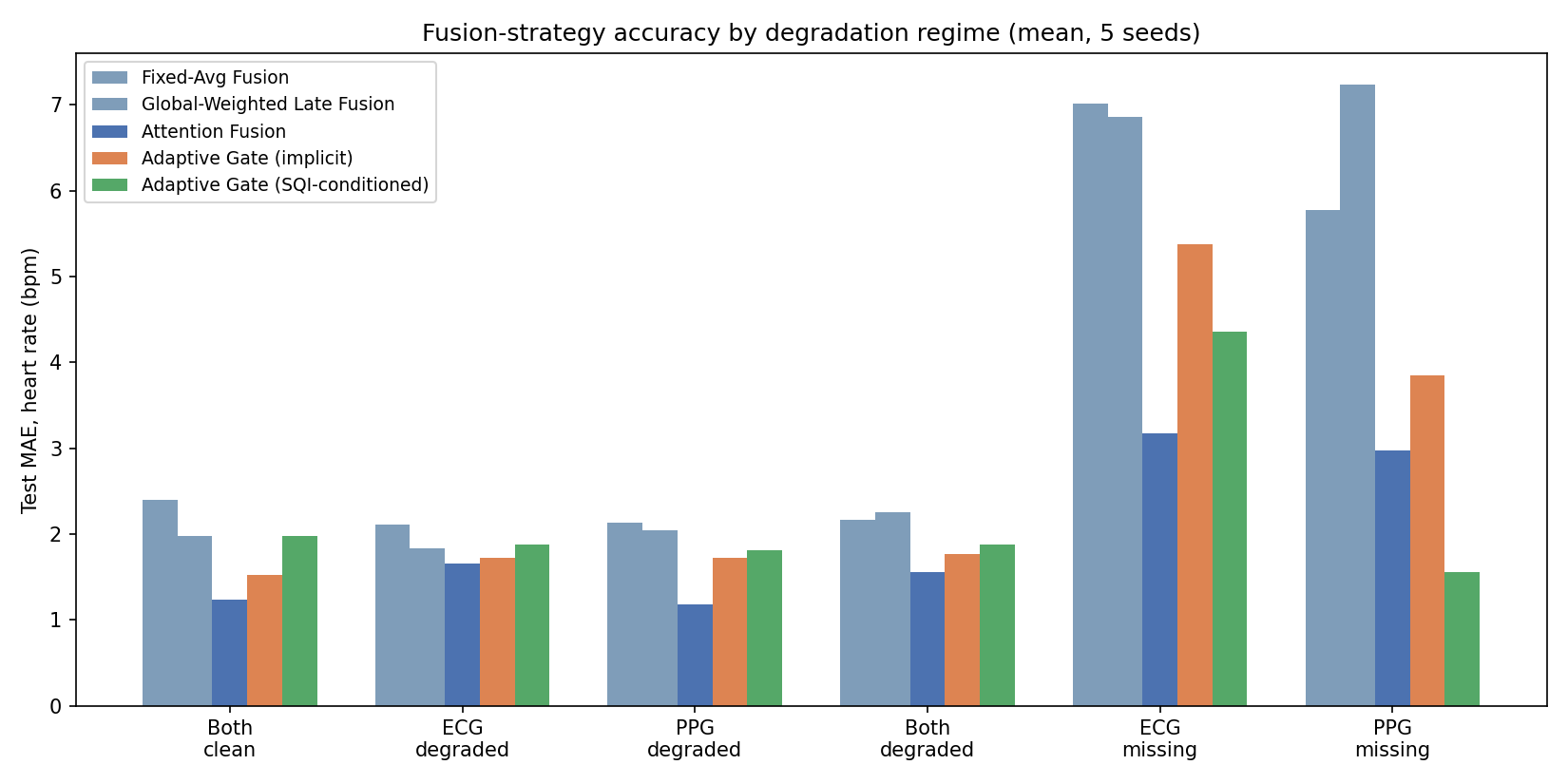}
\caption{Test MAE by degradation regime for five representative fusion strategies. Complete modality loss (right two groups) separates strategies far more sharply than graded degradation (left four groups).}
\label{fig:regimebreakdown}
\end{figure}

\subsection{Adaptive Gate Behavior and Effect of Explicit SQI Conditioning}
\label{sec:mechanistic}
\begin{figure}[t]
\centering
\includegraphics[width=\linewidth]{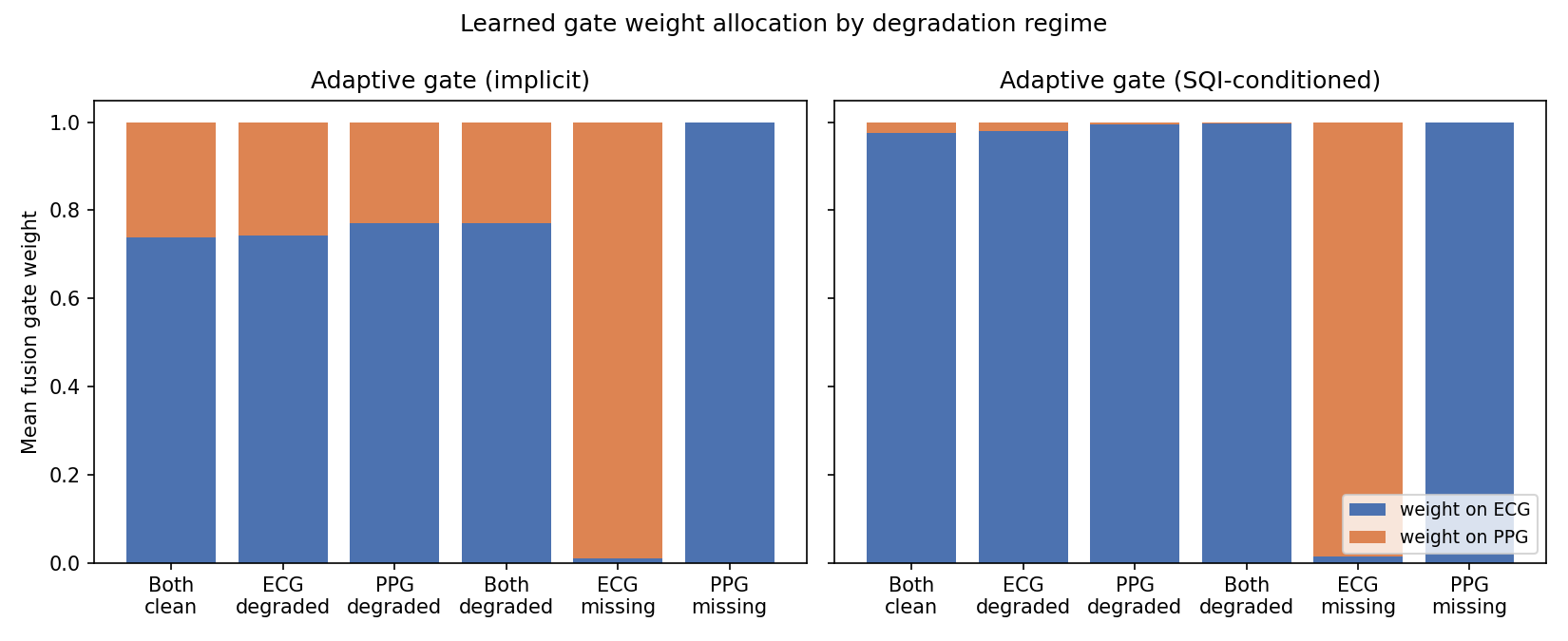}
\caption{Mean learned gate weight on ECG vs.\ PPG by degradation regime, for the implicit and SQI-conditioned adaptive gates. Both gates reallocate weight sharply only under complete modality loss (rightmost two regimes).}
\label{fig:gateweights}
\end{figure}

Fig.~\ref{fig:gateweights} shows the mean learned gate weight allocated to ECG, by regime, for both adaptive gates. Both gates allocate the large majority of weight to ECG when both modalities are present (implicit: 0.74--0.77 across clean and degraded-but-present regimes; SQI-conditioned: 0.975--0.998), and both correctly collapse to allocate $>98\%$ of weight to the surviving modality under complete dropout of the other (implicit: 0.009 with ECG missing, 1.000 with PPG missing; SQI-conditioned: 0.015 with ECG missing, 1.000 with PPG missing). Critically, however, \emph{neither} gate's weight allocation moves appreciably between the clean and severely-degraded-but-present conditions (implicit: 0.738 $\to$ 0.743; SQI-conditioned: 0.975 $\to$ 0.980, ECG-severity axis) -- both gates behave as an approximately step-like function of modality \emph{presence}, not a graded function of modality \emph{quality}.

We quantified this directly by correlating, per seed, the learned gate weight on ECG against the SQI difference $s_{\mathrm{ecg}}-s_{\mathrm{ppg}}$ across all test windows. Pooled across all regimes (including the missing-modality extremes), both gates show a strong association (implicit: mean Pearson $r=0.68$; SQI-conditioned: $r=0.70$, across seeds). Restricted to windows where \emph{both} modalities are present but graded-degraded -- the condition that actually determines whether a gate is quality-proportional rather than merely presence-detecting -- this association collapses (implicit: mean $r=0.24$; SQI-conditioned: mean $r=0.10$). That is, giving the gate direct numerical access to a validated signal-quality descriptor did not, in this experiment, teach it to use that information proportionally when both modalities remained present; the SQI-conditioned gate's advantage over the implicit gate is concentrated specifically in the missing-PPG condition (Section~\ref{sec:results}, Table~\ref{tab:mainresults}), not in graded-degradation responsiveness.

\subsection{Statistical Significance}
\label{sec:stats-results}
Table~\ref{tab:stats} summarizes the pre-registered, seed-level paired comparisons ($n=5$) with Holm correction across all 21 tests. No comparison survives correction. The two nominally closest comparisons (SQI-conditioned adaptive gate vs.\ attention fusion, on the ``both clean'' and ``PPG degraded only'' regimes) reached the minimum possible exact two-sided Wilcoxon $p$-value at $n=5$ ($p=0.0625$, i.e., a consistent direction across all five seeds) with large paired effect sizes (Cohen's $d_z=2.51$ and $1.50$, respectively), but neither survives Holm correction ($p_{\mathrm{Holm}}=1.0$). We report these as directionally consistent, large-effect-size, exploratory findings rather than confirmed differences, and we consider this an explicit limitation of a five-seed study rather than a null result to be minimized (Section~\ref{sec:limitations}).

\begin{table}[t]
\centering
\caption{Seed-level paired comparisons ($n=5$), selected regimes. Full 21-test family in supplementary material.}
\label{tab:stats}
\small
\begin{tabular}{@{}llrrrl@{}}
\toprule
Comparison & Regime & Mean diff. & $d_z$ & $p$ & $p_{\mathrm{Holm}}$ \\
\midrule
SQI vs.\ Attn. & both clean & +0.74 & 2.51 & 0.063 & 1.00 \\
SQI vs.\ Attn. & PPG degraded & +0.63 & 1.50 & 0.063 & 1.00 \\
SQI vs.\ Implicit & overall & $-$0.07 & 0.18 & 0.81 & 1.00 \\
Implicit vs.\ Attn. & overall & +0.40 & 0.56 & 0.31 & 1.00 \\
\bottomrule
\end{tabular}
\end{table}

\section{Discussion}
\label{sec:discussion}

\subsection{Why attention fusion performed best, with a specific exception}
Attention fusion's bidirectional cross-attention allows each modality's contribution to the fused representation to vary with the input, without requiring an explicit, hand-specified quality signal; unlike the gates in this study, it also retains \emph{both} attended representations (concatenated) rather than compressing to a single convex combination, giving it more representational capacity at the fusion step. This is consistent with its descriptively strongest overall performance. Its reversal under severe, single-modality ECG degradation (Section~\ref{sec:results}) indicates that this capacity does not translate into robustness against every degradation type, and that the mechanism most often recommended for its flexibility is not uniformly the most robust choice.

\subsection{Why adaptive gating did not track graded quality}
The gate architectures in this study learn from a 64-dimensional pooled embedding (optionally augmented with a 4-dimensional SQI vector) trained end to end against a single scalar heart-rate loss. A plausible explanation for the observed presence/quality asymmetry is that complete modality dropout produces a large, easily separable signal in embedding space (and an extreme SQI value), while graded degradation produces a comparatively subtle, continuous shift that a small gate network trained on a modest number of windows may simply not be incentivized to track, since the regression loss can often still be minimized by defaulting to whichever modality is globally more informative on average. This is consistent with -- though our design does not by itself establish causally -- a form of modality dominance during joint training \cite{wang2020}, discussed further below.

\subsection{Why SQI conditioning helped selectively}
Explicit SQI conditioning produced its clearest benefit exactly where the SQI signal itself is least ambiguous: complete PPG dropout collapses PPG SQI to near zero, a regime the gate can separate easily. Under graded degradation, where SQI changes are smaller and the underlying regression task may remain solvable without a large weight shift, the SQI-conditioned gate did not show materially different weight-allocation behavior from the implicit gate (Section~\ref{sec:mechanistic}). We regard this as the paper's central mechanistic finding: \emph{modality availability and modality quality are functionally distinct problems for adaptive fusion}, and providing a validated quality descriptor as an input feature does not, by itself, guarantee the network learns to use it proportionally.

\subsection{Modality competition as a plausible contributing factor}
Even when the SQI-conditioned gate correctly allocated $\sim$99\% of weight to PPG under ECG dropout, its accuracy (4.36 bpm) remained well short of a dedicated PPG-only model evaluated on the identical windows (1.47 bpm) -- despite the gate's decision being, in that regime, essentially correct. A plausible explanation is that because both gates favor ECG for the substantial majority of training windows (Section~\ref{sec:mechanistic}), the PPG-specific encoder receives systematically less gradient signal during joint training than it would in isolation, a pattern consistent with previously reported modality competition or modality dominance effects in multimodal network training \cite{wang2020}. We phrase this as a plausible contributing explanation rather than an established causal mechanism, since this study was not designed with the ablations (e.g., modality-balanced training curricula, gradient-magnitude monitoring) needed to establish it directly.

\section{Limitations}
\label{sec:limitations}
\begin{enumerate}
\item \emph{Five independent training seeds.} This provides adequate power to report consistent-direction trends and effect sizes but not confirmatory pairwise significance after multiplicity correction; the two largest observed effects ($d_z>1.5$) are promising candidates for confirmation in a higher-replicate follow-up.
\item \emph{Controlled, synthetic degradation.} The degradation model is physiologically grounded (Section~\ref{sec:degradation}) but is a controlled approximation of real-world corruption; absolute error magnitudes should not be read as real-world deployment performance, only as a comparative signal among architectures evaluated identically.
\item \emph{Modest dataset scale.} 792/234/324 train/validation/test window pairs, chosen for computational tractability, is small by deep-learning standards; per-regime test cells (as few as $\sim$18--90 windows) limit the precision of regime-specific estimates.
\item \emph{The gate and correlation analyses are descriptive, not causal.} They demonstrate consistent behavioral patterns but do not, by themselves, establish the underlying training-dynamics mechanism (Section~\ref{sec:discussion}).
\item \emph{Asymmetric strength of evidence.} The signal-processing front end (Section~\ref{sec:realdata}) is validated on real, independently sourced data; the comparative fusion-architecture findings (Sections~\ref{sec:results}--\ref{sec:mechanistic}) are not, and real-world robustness of the learned fusion models has not been established.
\end{enumerate}

\section{Conclusion}
We evaluated eight ECG--PPG fusion strategies under a controlled, physiologically-grounded degradation protocol spanning graded corruption and complete modality loss, building on a signal-processing front end independently validated against real ICU and fetal ECG recordings. Fusion-mechanism choice produced large, consistent differences in robustness. Attention fusion was the strongest general-purpose strategy but not uniformly so, failing specifically under severe single-modality ECG degradation. Both adaptive gating mechanisms correctly detected complete modality loss but showed little evidence of tracking graded signal quality when both modalities remained present; explicit conditioning on a validated signal-quality descriptor improved missing-modality robustness in one specific, asymmetric direction without resolving this broader limitation. We interpret modality availability and modality quality as distinct problems that adaptive multimodal physiological monitoring systems must address separately, and we report our statistical evidence, including its limits, explicitly rather than overstating what a five-seed comparative study can support.

\section*{Data and Code Availability}
The CardioFusion-AI signal-processing and fusion-architecture code is available at the project repository. Synthetic data generation code, trained-model evaluation scripts, and the full seed-level results tables underlying every reported statistic are available as supplementary material.

\end{document}